\documentclass[twocolumn,showpacs,aps,amsmath,amssymb]{revtex4}

\usepackage{epsfig,psfrag}
\usepackage{color}
\usepackage{graphicx}
\usepackage{dcolumn}
\usepackage{bm}
\usepackage{textcomp}
\usepackage{color}
\usepackage{natbib}
\DeclareGraphicsExtensions{.pdf,.png,.jpg,.eps}

\begin{document}

\title{Selective focusing of electrons and holes in a graphene-based superconducting lens}

\author{S. G\'{o}mez$^1$, P. Burset$^2$, W. J. Herrera$^1$ and A. Levy Yeyati$^2$}
\affiliation{
$^1$Departamento de F\'{\i}sica, Universidad Nacional de Colombia,
Bogot\'a, Colombia \\
$^2$Departamento de F{\'i}sica Te{\'o}rica de la Materia Condensada C-V,
Universidad Aut{\'o}noma de Madrid, E-28049 Madrid, Spain }

\date{\today}



\begin{abstract}

We show that a graphene {\it pnp} junction with a central superconducting electrode acts as a Veselago lens for incoming electrons by focusing them and their phase-conjugated counterpart (holes) into different points of the optical axis. This selective focusing suggested by a simple trajectory analysis is confirmed by fully microscopic calculations. Although the focusing pattern is degraded by deviations from the ideal conditions we show that it remains visible for a wide range of parameters. We discuss how this property can be useful for the detection of entangled electron pairs.
\end{abstract}

\maketitle

\section{Introduction.}
The possibility of fine-tuning the density of carriers in graphene together with their resemblance to massless particles like photons have made graphene a promising candidate for testing photonic analogies in electron transport.
Owing to the relativistic chiral nature of carriers there is a suppression of backscattering at a graphene-based {\it pn} junction known as Klein tunneling \cite{Katsnelson_2006}, which has been confirmed in recent experiments \cite{Stander_2009,Young_2009}. 
Furthermore, an electron beam flowing through a single {\it pn} junction experiences negative refraction \cite{Cheianov_2007} and thus this system has been proposed as the electronic equivalent of a Veselago lens \cite{Veselago_1968}. 
Thanks to this analogy, striking properties of the meta-materials such as perfect-lensing \cite{Pendry_perfect_lens} could be explored in graphene. 
In particular, supercollimation of electron beams \cite{Park_2008} has been proposed for graphene under periodic potentials and the focusing of electron beams has been studied in circular graphene {\it pn} junctions \cite{Cserti_2007}, graphene nanoribbons \cite{Xing_2010} and the surface of topological insulators \cite{Hassler_2010}. Graphene-based Veselago lenses have also been proposed as {\it filtering} systems for spin-polarized electron beams \cite{Zareyan_2010}. 
Additionally, advances in the construction of ballistic {\it pnp} junctions in graphene have been reported \cite{Ozyilmaz_2007,Gorbachev_2008,Velasco_2009,Chiu_2010}. 

New interesting possibilities can emerge if we consider the case of hybrid graphene-superconductor nanostructures \cite{Cuevas_2006,Ossipov_2007,Linder_2008}. Good contact can be achieved between lithographically defined superconducting electrodes and graphene layers \cite{Heersche_2007,Shailos_2007,Miao_2007,Andrei_2008}. In such devices a superconducting gap is induced by proximity effect on the graphene region underneath the metallic electrodes; in these conditions Andreev processes featuring conversion of electrons into holes take place \cite{Specular}. 
In a graphene-based normal-superconductor-normal (GSG) junction, local and crossed (CAR) Andreev reflections can occur if the width of the central superconducting electrode is comparable to the superconducting coherence length \cite{Cayssol_2008}. 
The time-reversal of these processes corresponds to the splitting of a Cooper pair from the superconductor into an entangled electron pair in the normal electrodes \cite{Feinberg_2000}. 
Although progress has been achieved in the experimental realization of Cooper pair splitters using carbon nanotube and semiconducting nanowire quantum dots \cite{Takis_2010,Basel_2010}, it is expected that graphene can provide even better conditions for the entanglement detection.

\begin{figure}
\epsfig{file=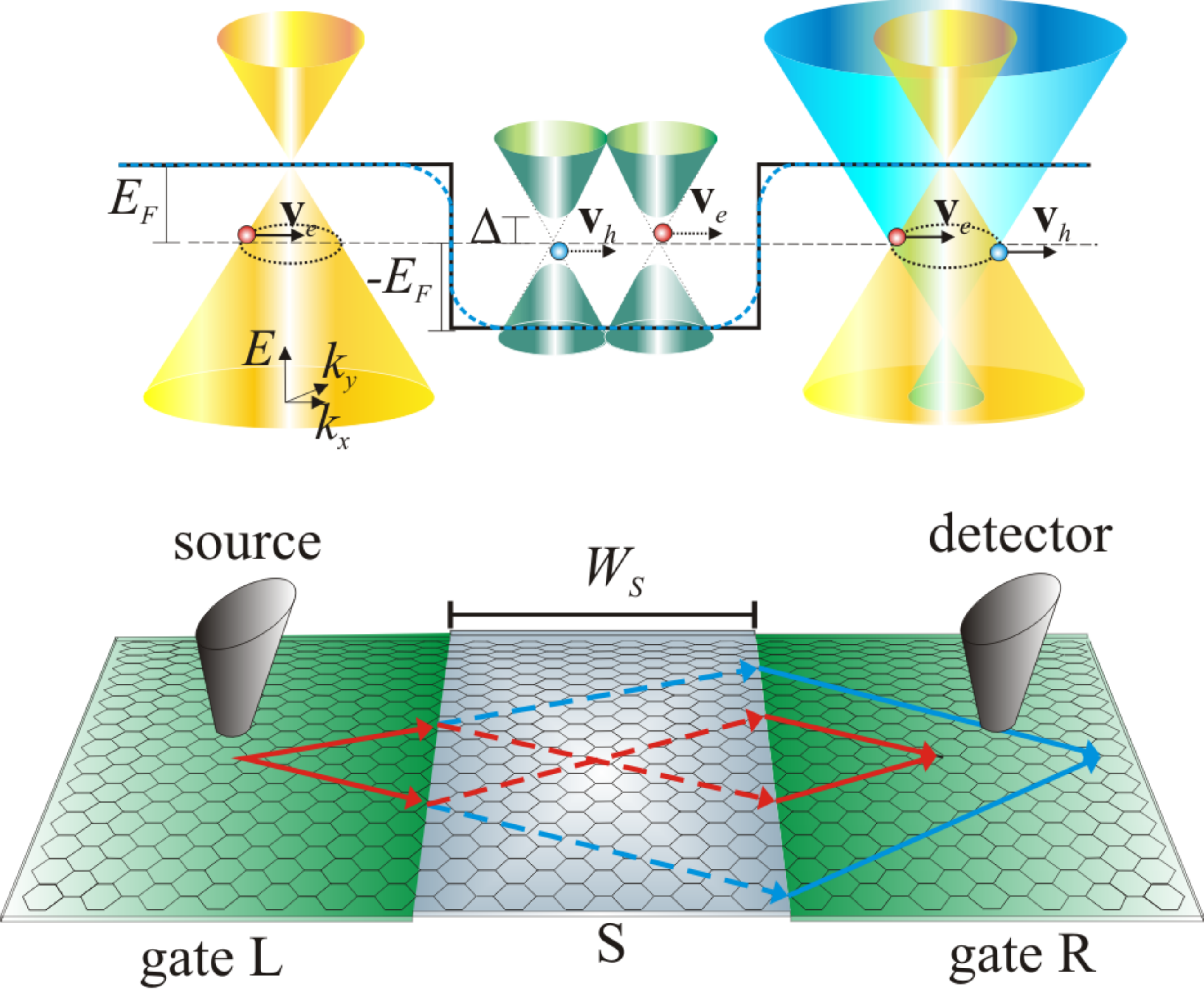,width=8.5cm}
\caption{
(Color online) 
Graphene superconducting {\it pnp} junction. The band structure of each region is shown on top for the case when the normal electrodes $L$ and $R$ are adjusted to have the opposite doping level than the superconducting central region $S$. Incoming electrons from the left electrode (solid red lines) transform into evanescent electron and hole-like excitations inside the superconductor (dashed red and blue lines, respectively). In the right region, electrons and holes (solid blue lines) focus into distinct regions.
}
\label{figure1}
\end{figure}

In this letter we propose to create a Veselago lens in a GSG junction to focus electrons and holes in different spatial regions. 
The idea is schematically depicted in Fig. \ref{figure1}. A graphene sheet is deposited on top of two independent gate electrodes and a central superconducting electrode (denoted $L$, $R$ and $S$ respectively). The superconducting electrode shifts the electronic bands of the underlying graphene region by transference of electrons to produce an {\it n}-doping effect. Subsequently, the doping level of each normal region is adjusted by the gate electrodes so the system behaves as a {\it pnp} junction but with the peculiarity that superconductivity is induced in the central region. Injection of electrons can be realized in the $L$ region by means of a local probe and when the central electrode is in the normal state the system acts as a Veselago lens focusing electrons at the region $R$. When superconductivity is ``switched on'', evanescent electron and hole-like states are created in the graphene region located under the central electrode.
In the right electrode, electrons and holes become propagating waves again and are focused into regions separated by hundreds of nanometers. This spatial separation would allow the detection of the transmitted holes by means of a second local probe (detector in Fig. \ref{figure1}). 

\begin{figure}
\epsfig{file=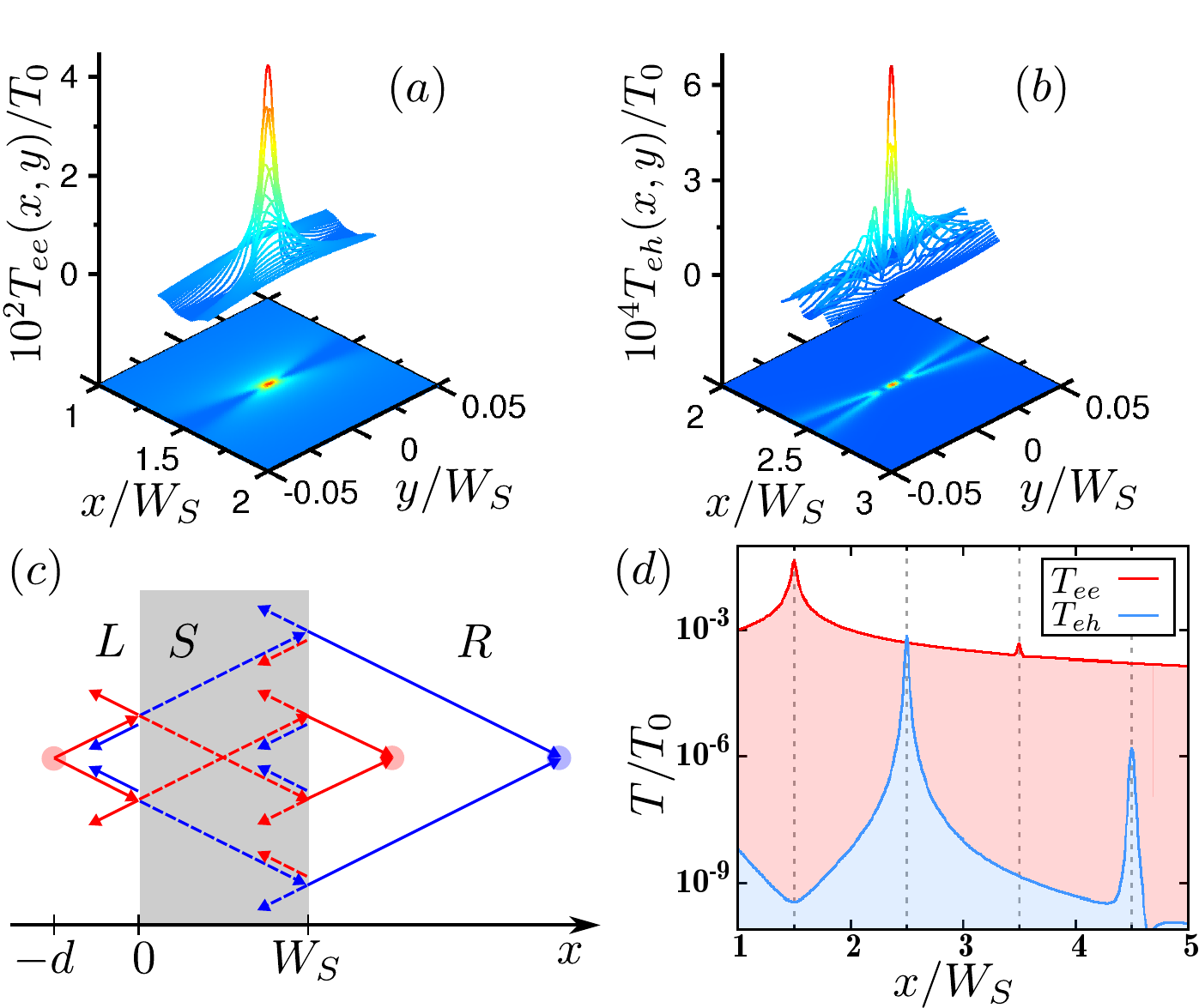,width=8.5cm}
\caption{
(Color online) 
(a) Map of the electron transmission $T_{ee}(x,y)$ near the focusing point $x=3W_S/2$. The width of the superconducting region is $W_S=\xi\sim 1\mu$m and the doping levels are $|E^{L,S,R}_F|=0.5$eV. 
(b) For the same parameters, map of the hole transmission $T_{eh}(x,y)$ near the focusing point $x=5W_S/2$.
(c) Sketch of the trajectories of electrons and holes through the superconductor into the region $R$. See the text for details.
(d) Logarithmic scale plot of the transmission of electrons (red) and holes (blue) into region $R$ for $y=0$, showing the profiles of (a) and (b) where the peaks due to internal reflections can be distinguished.
}
\label{figure2}
\end{figure}

\section{Independent focusing of electrons and holes.}
For modeling the Veselago lens depicted in Fig. \ref{figure1}, we consider an infinite plane of graphene with a superconducting electrode covering the region $0<x<W_S$, while the regions $x<0$ ($L$) and $x>W_S$ ($R$) remain in the normal state. 
Although graphene is not intrinsically superconducting, the superconducting electrode can induce a pairing amplitude $\Delta$ by proximity effect \cite{Tkachov_2007,Burset_2008,Black-schaffer_2008}. We choose the width of the central region to be comparable to the superconducting coherence length $W_{S}=\xi=\hbar v_F/\Delta$.
For a typical superconductor like Pb or Al, $\Delta \sim 1$meV and thus $\xi \sim 0.5 - 1 \mu$m.
An Al or Pd/Al electrode in the superconducting state induces a {\it n}-doping in the underlying graphene region estimated as $E_{S}^{F}\sim-0.5$eV \cite{Giovannetti_2008}. Since the gate potentials of the normal regions can be adjusted independently we choose them to be $E_{F}^{L,R}=-E_{F}^{S}$. This set of parameters define a {\it pnp} junction with a central superconducting electrode.
The transport properties between the two normal electrodes are computed in terms of one-particle Green functions following the method explained in Ref. \cite{Herrera_2010} and briefly presented in Appendix \ref{sec:app_a}. 
This allows to define an electron transmission probability $T_{ee}(x,y)$ from region $L$ to region $R$ (i.e. electron cotunneling EC), and a CAR probability $T_{eh}(x,y)$ where $(x,y)$ denotes the detector coordinates when the source coordinates are $(-d,0)$.
In Fig. \ref{figure2} we show the result of the microscopic calculation for the transmission probabilities. $T_{ee}$ exhibits a well-defined peak at $3d$ while we obtain a maximum of $T_{eh}$ at $5d$, as it is shown in Fig. \ref{figure2}(a) and Fig. \ref{figure2}(b) respectively. Both results are normalized to $T_0=T_{ee}(-d,0)$. The intensity of the electron focusing is much higher than that of the holes, i.e. $T_{ee}(3d,0)\sim 100T_{eh}(5d,0)$, thus indicating that the EC signal is much greater than the CAR signal. However, the spatial separation between peaks is exactly the width of the superconducting region, which is of the order $1\mu$m. This would allow to detect each signal independently as we discuss in more detail below.

These results can be qualitatively explained with a simple analysis of the group velocities of the particles at each region (see Fig. \ref{figure2}(c)).
For a perfectly symmetric {\it pnp} junction, the normal regions $L$ and $R$ are {\it p}-doped with $E_F>0$, while the superconducting region $S$ is n-doped with $-E_F$. 
In the heavily doped regime $|E_F|\gg \Delta, E$ an analysis of the propagation of waves based on classical trajectories is sensible since $\lambda_F \ll W_S$. Even for the evanescent waves within the superconducting region and with $|E|<\Delta$ this type of analysis is valid considering waves with $k_F^S\sim k_F^{L,R} \sim E_F/\hbar v_F$. Subsequently, the angle of incidence and transmission of particles is defined as $\phi(k_y,E_F)=\pm\arcsin[\hbar v_F k_y/ E_F]$, where the sign depends of the doping level of each region. As a result, the group velocities of electrons and holes can be written as $\mathbf{V}_{e,h}=\pm\varepsilon v_F \mathbf{k}_{e,h}/|\mathbf{k}_{e,h}| = \pm\varepsilon v_F \left( \cos{\phi}, \sin{\phi} \right)$ where $\varepsilon=1(-1)$ for quasiparticles in the conduction (valence) band. 
From the conservation of the component of the wave vector parallel to the interface we reach the electronic equivalent of Snell's law at each interface \cite{Cheianov_2007,Hassler_2010} which allows to define a relative refraction index at each interface $n_{LS,SR} = \sin{\phi^{L,S}}/\sin{\phi^{S,R}} = -\varepsilon$. Taking into account the sign due to the band index (conduction or valence) and the one due to particle index (electron or hole), when the particle type is the same at both sides of the interface ($\varepsilon=1$), the change of band causes a negative refraction with $n=-1$. On the other hand, when the particle type is not conserved ($\varepsilon=-1$), there is no negative refraction and $n=1$. 
This explains the classical trajectories sketched in Fig. \ref{figure1} and Fig. \ref{figure2}(c): incoming electrons from the region $L$ (solid red lines) transform into electron and hole-like excitations inside the superconductor (dashed red and blue lines, respectively). 
While the former experiences a negative refraction because the particle type is conserved, the latter follows the same path as the incoming electron. 
At the second interface, the processes that preserve the particle type experience a negative refraction and are focused to form an image in the optical axis $x$ (for an analysis of the transmission amplitudes see Appendix \ref{sec:app_b}).

Furthermore, at each interface {\it specular} reflection occurs when the particle type is conserved while {\it retro}-reflection happens otherwise (Fig. \ref{figure2}(c)). Electron and hole-like excitations can endure two consecutive specular reflections inside the superconductor to create a new electron or hole beam in the normal region $R$. This leads to a sequence of alternated electron and hole focusing points at the optical axis. In Fig. \ref{figure2}(d) we show the microscopic calculation in which we obtain peaks of $T_{ee}$ at $x=(2m-\frac{1}{2})W_S$ and peaks of $T_{eh}$ at $x=(2m+\frac{1}{2})W_S$, with $m=1,2,...$. The intensity of these peaks decays exponentially with the distance to the superconductor, consistently with the behavior of the proximity effect in a graphene-superconductor interface \cite{Tkachov_2007,Burset_2008,Black-schaffer_2008}.  

\begin{figure}
\epsfig{file=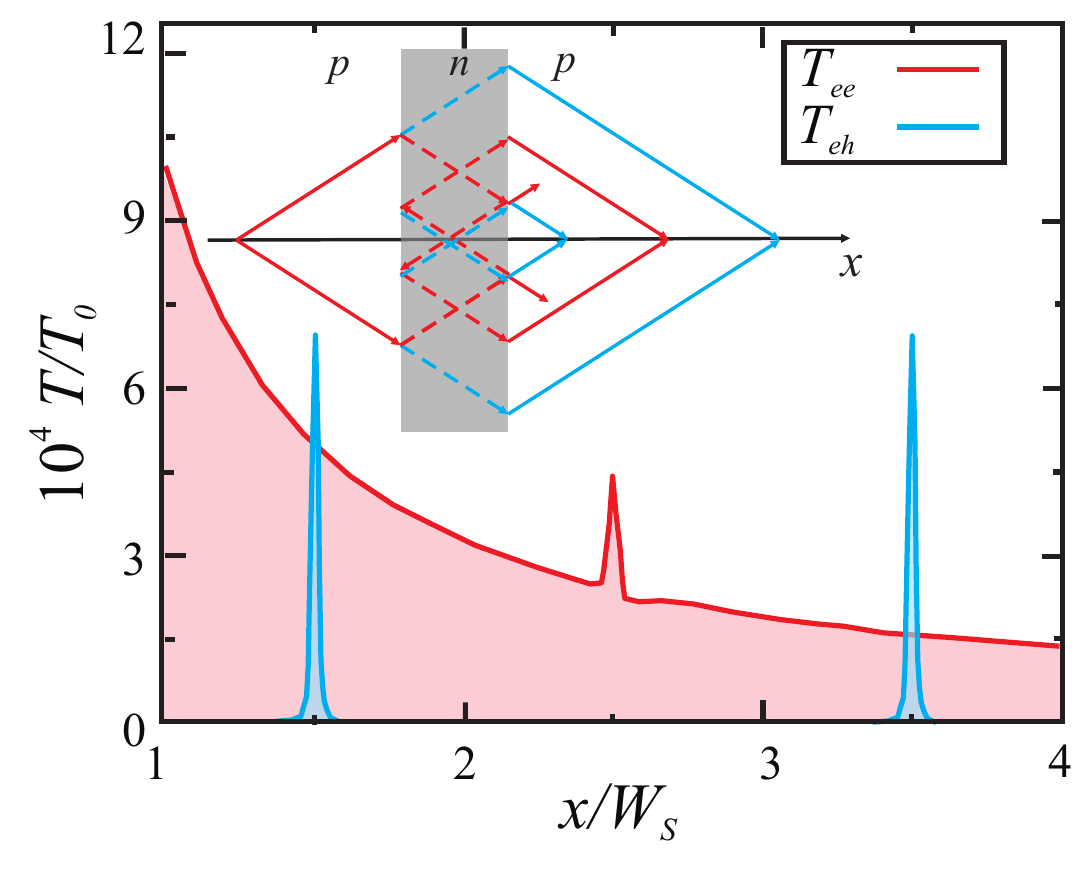,width=8.5cm}
\caption{
(Color online) 
Microscopic calculation for $T_{ee}(x,0)$ and $T_{eh}(x,0)$ when the injection of electrons is done at $d=-3W_S/2$. The inset shows a sketch of the classical trajectories. Notice that peak of $T_{eh}$ at $x=3.5W_S$ is $\sim 5$ times larger than the EC background.
}
\label{figure3}
\end{figure}

While the previous analysis explains the separate focusing of electrons and holes, it would be desirable that the background EC conductance at the CAR peak could be further reduced. This can be achieved by increasing the injection distance $d$. Indeed, the injection point $(-d,0)$ determines the origin of the sequence of focusing points in which the separation between maxima of $T_{ee}$ and that of $T_{eh}$ is $W_S$. For $d<W_S$ the sequence of points starts always with a maximum of $T_{ee}$. On the other hand, for $d>W_S$, electrons are only transmitted into the region $R$ after two or more internal specular reflections. 
However, a combination of one specular reflection and one retro-reflection, which changes the particle type, allows to have a hole focusing point at $x=1.5W_S$, i.e. to the left of the electron focusing point at $x=2.5W_S$ (see Fig. \ref{figure3} for a sketch of the trajectories and the microscopic calculation). The intensity of this peak is the same as the CAR peak appearing at $x=3.5W_S$, to the right of the electron focusing point, which comes from a transmitted hole without internal reflections. Both peaks of $T_{eh}$ in Fig. \ref{figure3} are greater than the peak of $T_{ee}$ and in particular the peak at $x=3.5W_S$ is $\sim 5$ times larger than the background EC contribution.

\begin{figure}
\epsfig{file=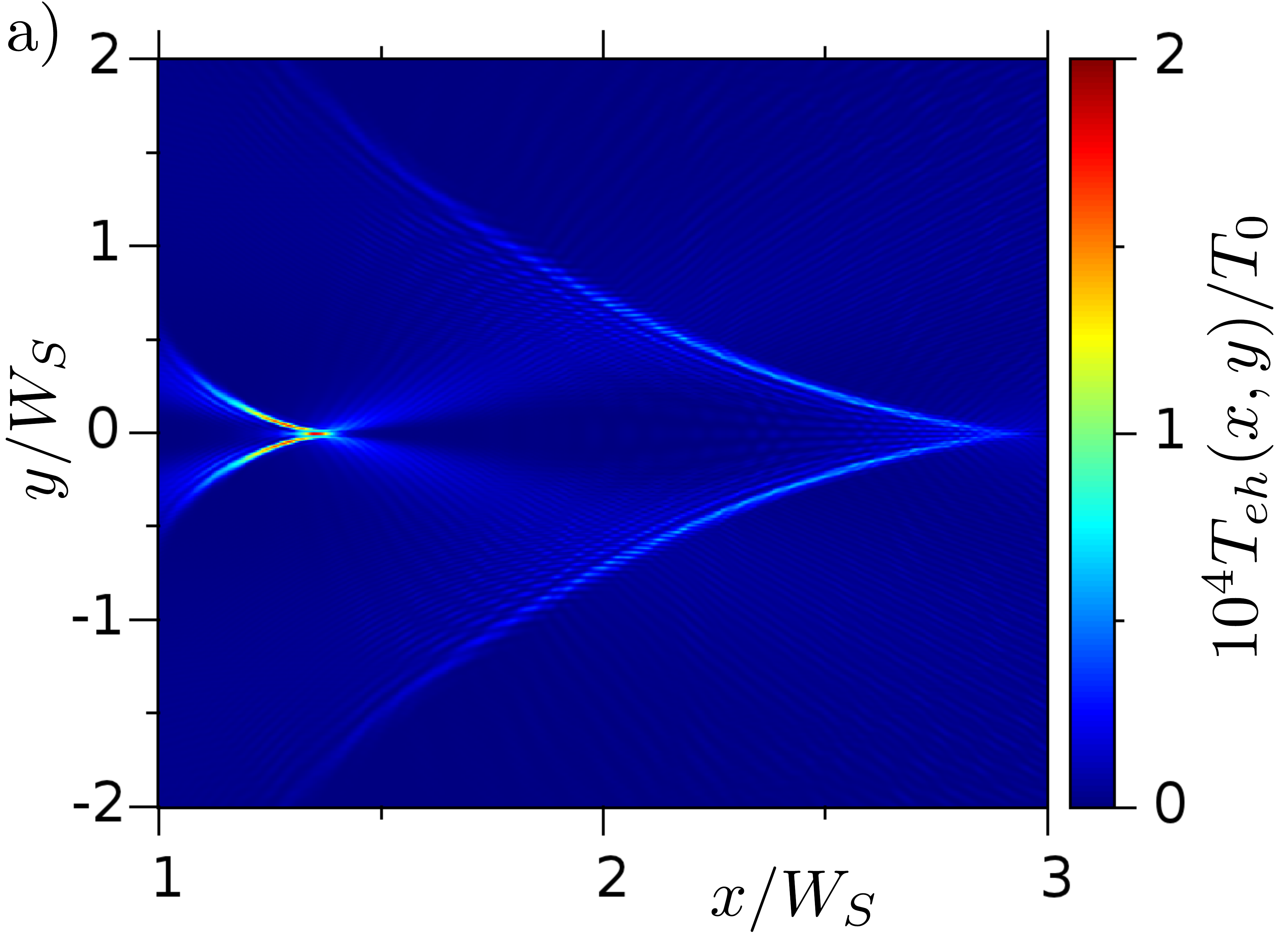,width=8.5cm}
\epsfig{file=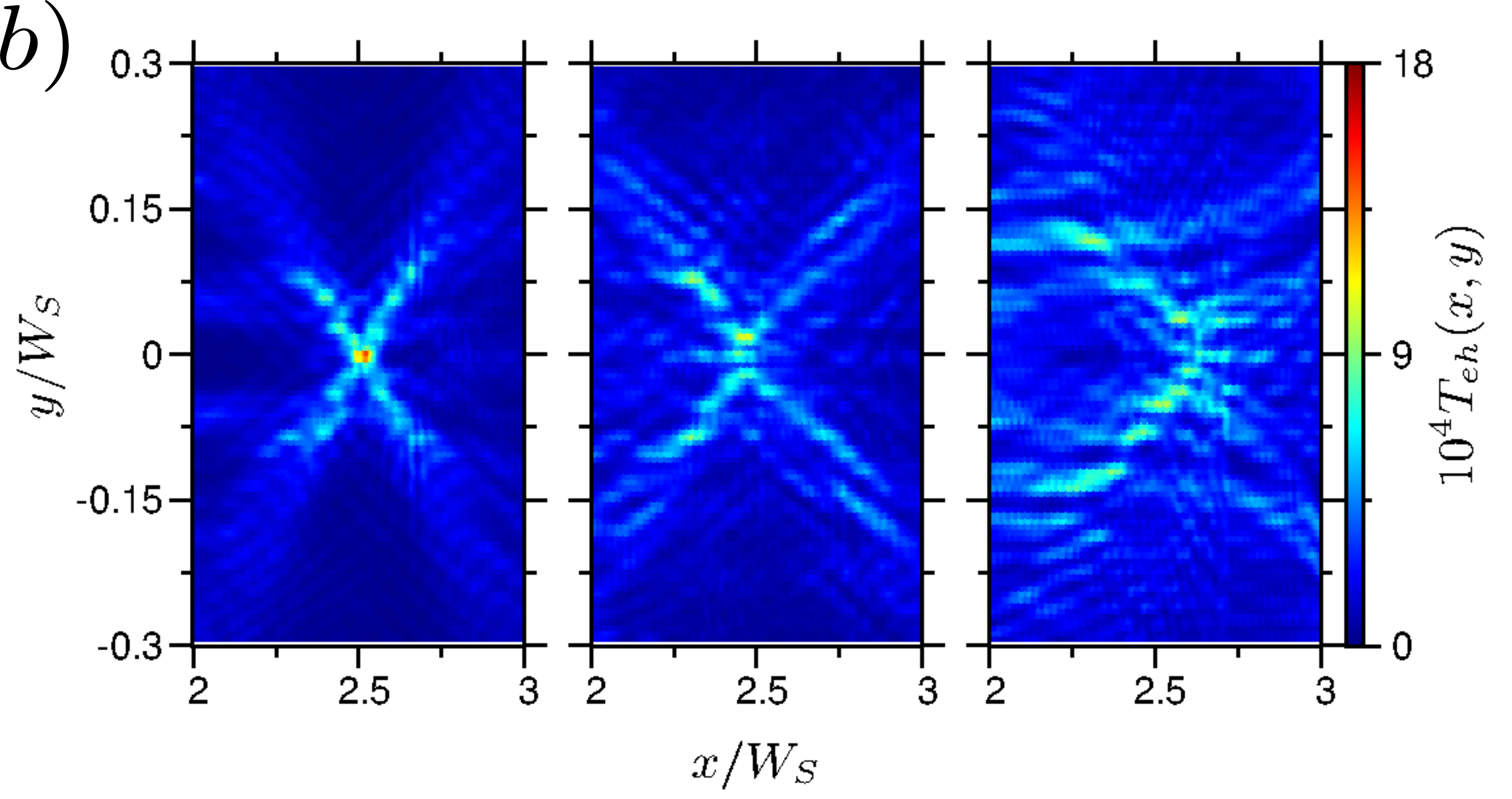,width=8.5cm}
\caption{
(Color online) 
(a): Map of $T_{eh}(x,y)$ for an asymmetric \textit{pnp} junction. We set $E_F^R=0.4$eV and keep the rest of the parameters the same as in Fig. \ref{figure3}, where $d>W_S$. The refracted waves in the right region form a characteristic interference pattern showing two caustic curves near the cusp for each of the transmitted holes. 
(b): Map of $T_{eh}(x,y)$ for a \textit{pnp} junction calculated using the TB model. The length of the central superconducting region is $W_S=85.2$nm. The gate potentials used are $E_F^{L,R}=-E_F^S=0.6$eV, with $\Delta=2.7$meV. The left panel includes no disorder. The central panel has $V_0=54$meV. The right panel has $V_0=81$meV. The smearing of the potential for the three panels has a range of $\sim 35$nm.
}
\label{figure4}
\end{figure}

\section{Deviations from the ideal case.}
We have considered thus far that each interface is a perfectly symmetric {\it pn} junction, i.e. $E_F^p=-E_F^n$. 
If the doping level of one of the regions is not perfectly aligned with the next we have that $n\neq-1$ for that interface. In Fig. \ref{figure4}(a) we show a map of $T_{eh}(x,y)$ with $E_{F}^{L}=-E_{F}^{S}=0.5$eV and $E_{F}^{R}=0.4$eV where the refraction index of the SR interface becomes $n=-E_{F}^{R}/E_{F}^{L}=-0.8$. For this case the classical trajectories are deformed at the region $R$. The focal points are displaced from the ones shown in Fig. \ref{figure3} and the envelope of the refracted rays becomes a caustic curve \cite{Shendeleva_2008}. The focusing pattern is reproduced but the intensity of the cusps is not the same, as it was the case for the symmetric junction. 
Caustic curves for electrons in asymmetric {\it pn} junctions have been predicted to appear \cite{Cheianov_2007,Cserti_2007,Hassler_2010} but we show here that these curves appear in spite of being originated from hole-like excitations inside the superconductor. We thus conclude that a doping imbalance between regions has the same effect on the CAR signal that on the EC one.

The results presented thus far correspond to an infinite layer of pristine graphene under a perfectly sharp potential profile. A more realistic model should include size effects such as a graphene layer with a finite length, a potential profile varying smoothly along the sample and the inclusion of electron-density inhomogeneities (i.e charge puddles \cite{puddles_2008,puddles_2009}). Using a tight-binding (TB) model we explore the stability of the focusing pattern under these premises. 
We define a defect-free graphene strip of total length $W\sim 900$nm, with a central superconducting region of length $W_S\sim 90$nm and coupled to normal metallic electrodes at the edges (the details of the TB model are presented in the Appendix \ref{sec:app_c}). 
A smearing of the potential profile within a range of $30-40$nm is introduced at each normal-superconductor interface. In addition, we introduce random inhomogeneities of the potential profile of strength $V_0$ over an area of typical length $d\sim 20-30$nm.

We show in the left panel of Fig. \ref{figure4}(b) the TB results for $T_{eh}(x,y)$ in the absence of disorder ($V_0=0$). The focusing spot at $x=2.5W_S$ is clearly distinguishable although some diffraction effects are present due to the finite length of the system and the smearing of the potential. When we introduce disorder of strength $V_0=54$meV (central panel) the intensity of the focusing spot is reduced but its size remains almost unchanged. In the right panel, when the disorder strength is increased to $V_0\sim 81$meV, diffraction effects overcome the focusing pattern. The smearing of the potential has a range of $\sim 35$nm for the three panels. As it was demonstrated in Ref. \cite{Xing_2010} for a smooth \textit{pn} junction, the smearing of the potential introduces a small diffraction effect in the transmission, reducing the intensity of the focusing point, but leaving the extension of the spot almost unchanged. We also find that the focusing pattern is robust against disorder caused by charge puddles of width $\sim 20-30$nm and strength $V_0 \lesssim 80$meV. 
These parameters are well above the measured inhomogeneities in graphene, which are bounded to $30$meV over $20-30$nm \cite{puddles_2008,puddles_2009}.

\section{Conclusions.}
In conclusion we have shown that selective focusing of electrons and holes can be produced in a GSG junction.
In addition, the geometry can be tuned in order that the CAR peak dominates over the EC background. 
We have also shown that the focusing is robust against deviations from the ideal conditions. 
Under these premises, a possible experimental realization of this proposal is sketched in Fig. \ref{figure1}, where electrons are injected at the source electrode and collected at the detector electrode. If the source is a fixed electrode, a mobile detector would be able to distinguish between the EC signal and the CAR signal by moving from one focusing point to the other. On the other hand, if the detector is fixed, by moving the source electrode the focusing pattern can be adjusted to reach the fixed electrode. Although the nonlocal electron-hole transmission is reduced by a factor $10^{-4}$ with respect to the transmission at the injection point, the total nonlocal conductance can reach a measurable value when adding the contribution of many channels. 
These properties open an interesting route for the detection of entangled electron pairs over distances of $\sim 1\mu$m.

The authors would like to thank S. Csonka and P. Recher for fruitful discussions.
This work was supported by COLCIENCIAS, project 110152128235 (SG and WJH) and  MICINN-Spain via grant FIS2008-04209 and EU project SE2ND (PB and ALY).

\appendix

\section{Modeling the system.}
\label{sec:app_a}
We consider an impurity-free graphene sheet in the $x-y$ plane. A superconducting electrode is deposited on top of the region $0<x<W_S$. The low-energy excitations of the system are described by the Dirac-Bogoliubov-de Gennes (DBdG) equations
\begin{equation}
\left( \begin{array}{cc}
H-V(x) & \Delta(x)  \\ 
\Delta(x)  & V(x)-H
\end{array}
\right)\! \left( 
\begin{array}{c}
u \\ 
v
\end{array}
\right)\! =\! E\left( 
\begin{array}{c}
u \\ 
v
\end{array}
\right)  ,
\label{eq:DBDG}
\end{equation}
where $H=\hbar v_F \left( \hat{\sigma}_x k + \hat{\sigma}_y q\right)$ is the one particle Dirac Hamiltonian with Fermi velocity $v_F$, $\Delta(x)$ is the pairing amplitude, $V(x)$ is the potential profile and $E>0$ is the excitation energy. We impose rigid boundary conditions at the normal-superconducting interfaces to the pairing and the electrostatic potentials such that $\Delta(|x|>W_S)=0$,  $\Delta(|x|<W_S)=\Delta$ and $V(x<0)=E_F^L$, $V(0<x<W_S)=E_F^S$, $V(x>W_S)=E_F^R$. Whenever the pairing potential is assumed constant and non-zero, the low energy spectrum is given by $E=\sqrt{\Delta^2+\left( E_F^S - \hbar v_F \sqrt{k^2+q^2} \right)^2}$. We define the
transversal momentum as $\hbar v_F k_{\pm}= \sqrt{\left( E_F^S \pm \Omega \right)^2 - q^2}$, with $\Omega=\sqrt{E^2-\Delta^2}$ and $\hbar q$ the conserved momentum parallel to the interfaces. The pairing potential couples electrons and holes from different valleys. Eq. \ref{eq:DBDG} is therefore written in Nambu and pseudospin space, omitting the valley and spin degeneracies.

The transport properties can be expressed in terms of one-particle Green functions which satisfy $\left[ (E\pm i0^+) - {\cal H}(x) \right]G^{r,a}(x,x')=\delta(x-x')$, where ${\cal H}$ denotes the full Hamiltonian of the left hand side of Eq. \ref{eq:DBDG}. We calculate the Green functions by solving separately each region and combining the results following the method explained in Ref. \cite{Herrera_2010}. To fully resolve spatially the Green functions we use the Fourier transform ${\cal G} (x,x';y-y')=\int \mathrm{d}q e^{iq(y-y')} G(x,x')$. Therefore, by setting the electron injection in the left region $(x'<0)$ we analyze the electron transmission into the right region defining $T_{ee} \propto \mathrm{Tr}\left|{\cal G}_{ee}(x,x';y-y')\right|^2$ and the CAR probability as $T_{eh} \propto \mathrm{Tr}\left|{\cal G}_{eh}(x,x';y-y')\right|^2$, where the trace is done in the pseudospin space.

\section{Scattering amplitudes at the graphene-superconductor interface.}
\label{sec:app_b}
We consider a normal-superconductor interface along the $y$-direction in a graphene sheet, with the normal region extended at $x<0$. An incoming electron into the interface from the normal region can be reflected as an electron or a hole with probability amplitudes $r_{ee}$ and $r_{eh}$ respectively or it can be transmitted into the superconducting region as an electron-like or a hole-like quasiparticle, with probability amplitudes $t_{ee}$ and $t_{eh}$ respectively. The scattering states in both regions are
\begin{eqnarray}
\psi^L(x) \!\!&=&\!\! e^{ik^L_e x}\!\! \left(\!\! \begin{array}{c}
\varphi_{1e}^L \\ 0 \end{array} \!\!\right) \!+\!
r_{ee} e^{-ik^L_e x}\!\! \left(\!\! \begin{array}{c}
\varphi_{2e}^L \\ 0 \end{array} \!\!\right) \nonumber \\
&& \!+ 
r_{eh} e^{ik^L_h x}\!\! \left(\!\! \begin{array}{c}
0 \\ \varphi_{1h}^L \end{array} \!\!\right) , \\
\psi^S(x) \!\!&=&\!\! t_{ee} e^{ik^S_e x}\!\! \left(\!\! \begin{array}{c}
u \varphi_{1e}^S \\ v \varphi_{1e}^S \end{array} \!\!\right) \!\!+\!\! 
t_{eh} e^{-ik^S_h x}\!\! \left(\!\! \begin{array}{c}
v \varphi_{2h}^S \\ u \varphi_{2h}^S \end{array} \!\!\right) ,
\label{eq:scatt_states}
\end{eqnarray}
where the dependence on $y$ has been omitted because the vertical momentum $q$ is conserved. Following the notation explained in Ref. \cite{Herrera_2010}, we have defined the bispinors in sublattice space $\varphi^T_{1e,1h}=(1,e^{\pm i\alpha}_{e,h})$, $\varphi^T_{2e,2h}=(1,-e^{\mp i\alpha_{e,h}})$ and the BCS coherence factors $u^2(v^2)=(E \pm \Omega)/2\Delta$, which are normalized so that $|u|^2+|v|^2=1$. We are interested in the heavily doped regime with $|E_F^{L}|=|E_F^S|\equiv E_F \gg \Delta,E$, which satisfies the mean-field approach for superconductivity. In this regime $k_{e,h}^L \approx k_{e,h}^S$ and $k_e^{L,S}=k_h^{L,S}$. The angles at each region are thus defined as $\alpha_e^L=\alpha_h^L=\phi$ and $\alpha_e^S=\alpha_h^S=s\phi$ with $\phi = \arcsin{\hbar v_F/E_F}$ and $s=\mathrm{sign}(E_F^L)\mathrm{sign}(E_F^S)$, which is positive when there is no change in the doping level at the interface and negative otherwise (i.e. for a {\it np} or a {\it pn} junction). Matching the scattering states at the interface it is straightforward to obtain the reflection and transmission amplitudes (a detailed description is given in Ref. \cite{Linder_2008}), in the energy regime of this work they reduce to
\begin{eqnarray}
r_{ee} \!&=&\! \frac{(e^{i\phi}+e^{is\phi})(e^{-i\phi}+e^{-is\phi})} {e^{-2i\phi}+e^{-2is\phi}} , \\
r_{eh} \!&=&\! \frac{-2i\cos{\phi}e^{-is\phi}} {e^{-2i\phi}+e^{-2is\phi}} , \\
t_{ee} \!&=&\! (1-i)\cos{\phi} \frac{e^{-i\phi}+e^{-is\phi}} {e^{-2i\phi}+e^{-2is\phi}} , \\
t_{eh}\!&=&\! -(1+i)\cos{\phi} \frac{e^{-i\phi}-e^{-is\phi}} {e^{-2i\phi}+e^{-2is\phi}} .
\label{eq:amplitudes}
\end{eqnarray}
When the NS interface is equivalent to a perfectly transparent symmetric {\it nn} junction the incoming electron is transmitted into the superconductor only as an electron-like quasiparticle since $s=1$ and thus  $t_{eh}=0$. 

On the other hand, we can analogously examine the reflection and transmission amplitudes of an incoming electron-like excitation from the superconductor into the rightmost normal region. The scattering states are thus
\begin{eqnarray}
\psi^S(x) \!\!&=&\!\! e^{ik^S_e x}\!\! \left(\!\! \begin{array}{c}
u \varphi_{1e}^S \\ v \varphi_{1e}^S \end{array} \!\!\right) \!\!+\!\! 
r'_{ee} e^{-ik^S_e x}\!\! \left(\!\! \begin{array}{c}
u \varphi_{2e}^S \\ v \varphi_{2e}^S \end{array} \!\!\right) \nonumber \\
&& \!+ r'_{eh} e^{ik^S_h x}\!\! \left(\!\! \begin{array}{c}
v \varphi_{1h}^S \\ u \varphi_{1h}^S \end{array} \!\!\right) , \\
\psi^R(x) \!\!&=&\!\! t'_{ee} e^{ik^R_e x}\!\! \left(\!\! \begin{array}{c}
\varphi_{1e}^R \\ 0 \end{array} \!\!\right) \!+\!
t'_{eh} e^{-ik^R_h x}\!\! \left(\!\! \begin{array}{c}
0 \\ \varphi_{2h}^R \end{array} \!\!\right) .
\label{eq:scatt_states_SN}
\end{eqnarray}
The resulting amplitudes, in the energy regime used are
\begin{eqnarray}
r'_{ee} \!&=&\! \frac{1-e^{2i\phi}} {e^{-2i\phi}+e^{-2is\phi}} , \\
r'_{eh} \!&=&\! ie^{i\phi} \frac{(e^{is\phi}+e^{-is\phi})} {e^{-2i\phi}+e^{-2is\phi}} , \\
t'_{ee} \!&=&\! \frac{(1+i)}{2} \frac{1+e^{-2is\phi}+e^{i\phi}(e^{is\phi}+e^{-is\phi})} {e^{-2i\phi}+e^{-2is\phi}} , \\
t'_{eh}\!&=&\! \frac{(1-i)}{2} \frac{1+e^{2is\phi}-e^{i\phi}(e^{is\phi}+e^{-is\phi})} {e^{-2i\phi}+e^{-2is\phi}} .
\label{eq:amplitudes_SN}
\end{eqnarray}
The incoming electron-like excitation can be reflected inside the superconductor both preserving and changing the particle type. The former is a specular reflection with a change of sign in both components of the velocity while the latter is a retro-reflection in which the new excitation follows back the path of the incident one. Equivalent results are obtained for an incoming hole-like excitation.
As in the previous case, for a {\it nn} junction ($s=1$) there is only transmission into the normal region preserving the particle type since $t'_{eh}=0$.

To summarize, for a {\it pnp} junction the incident electron from the normal region $L$ splits into electron and hole-like excitations inside the superconductor and this splitting determines the focusing points of each type on the normal region $R$. On the other hand, for a {\it nnp} junction only electron-like excitations are created inside the superconductor and thus there is only focusing of electrons in region $R$. This condition can be relaxed if the transparency of the interface is not perfect or if the junction is not symmetric (i.e. $|E_F^N|\neq |E_F^S|$). 

\section{Tight-binding model.}
\label{sec:app_c}
In order to describe a more realistic system, we analyze the electronic states of a defect-free graphene layer using the tight-binding approximation,
\begin{equation}
\hat{H}
= - t_g \sum_{<ij>} \hat{c}^{\dagger}_{i} \hat{c}_{j}
+ \sum_{i} V_{i} \hat{c}^{\dagger}_{i}\hat{c}_{i}
\, \, ,
\end{equation}
where $t_g=2\hbar v_F/3a_0\approx 2.6$eV denotes the hopping element between nearest carbon atoms on the hexagonal lattice, $a_0\approx 0.14$nm is the smallest carbon-carbon distance and $V_i$ is the potential applied to the lattice. The spin degree of freedom has been omitted due to degeneracy. We assume a finite horizontal length $W=W_L+W_S+W_R$, where $W_{L,R}$ are the lengths of the normal regions and $W_S$ is the length of the superconducting one. We have well-defined zigzag edges along the vertical direction. 
We impose periodic boundary conditions in the direction parallel to the edges in order to work in the regime in which the vertical distance is much larger than the horizontal one and thus have a continuum of transversal modes.

To compute numerically the transport properties we connect the zigzag edges to heavily doped graphene leads, maintaining the graphene sublattice structure at the edges and thus representing the experimental situation in which the electrodes are deposited on top of the graphene layer \cite{Schomerus_2007,Blanter_2007,Brey_2007c,Burset_2008,Burset_2009,Burset_2011}. 
The self-energies on the graphene sites at the layer edges are thus given by $\Gamma_L=\Gamma_R=t_g$. Analogously, the central superconducting region has an effective coupling with the superconducting electrode given by $\Gamma_S \sim \Delta\sim 1$meV.

The electrostatic potential profile along the graphene strip is defined as
\begin{eqnarray}
\!\!V(x)\! &\!=\!&\! E_{F}^{S}+\frac{E_{F}^{L} \!-\! E_{F}^{S}}{\pi}\left[ \frac{\pi}{2}-\arctan{\frac{x-W_L}{\alpha a_0}}\right] \nonumber \\
&+& \frac{E_{F}^{R} \!-\! E_{F}^{S}}{\pi}\left[ \frac{\pi}{2}\!+\!\arctan{\frac{x\!-\!W_L\!-\!W_S}{\alpha a_0}}\right] ,
\label{eq:pot2}
\end{eqnarray}
where $E_F^{L,R}$ are the gate potentials of the normal regions, $E_F^{S}$ is the doping level of the superconducting region and the parameter $\alpha$ controls the smearing of the potential at the interfaces. In addition, we introduce an extra term $\delta V$ in the potential. This term accounts for disorder in the distribution of charge on the graphene strip. $\delta V$ takes random values in the range $[-V_0,V_0]$ over an area of typical size $d$.

\end{document}